\documentclass[apj]{emulateapj}
\usepackage{graphicx}
\usepackage{subfigure}
\usepackage {natbib}

\begin{document}

\title{The Mass-Metallicity Relation Of A $Z\sim2$ Protocluster With MOSFIRE}

\author{\sc Kristin R. Kulas\altaffilmark{1,*}, Ian S. McLean\altaffilmark{1}, Alice E. Shapley\altaffilmark{1}, Charles C. Steidel\altaffilmark{2}, Nicholas P. Konidaris\altaffilmark{2}, Keith Matthews\altaffilmark{2}, Gregory N. Mace\altaffilmark{1}, Gwen C. Rudie\altaffilmark{2}, Ryan F. Trainor\altaffilmark{2}, Naveen A. Reddy\altaffilmark{3}}
\altaffiltext{*}{Email: kkulas@astro.ucla.edu}
\altaffiltext{1}{Department of Astronomy, University of California, Los Angeles, 430 Portola Plaza, Los Angeles, CA 90095,  USA}
\altaffiltext{2}{California Institute of Technology, MS 249-17, 1200 East California Boulevard, Pasadena, CA 91125, USA}
\altaffiltext{3}{Department of Physics and Astronomy, University of California, Riverside, 900 University Avenue, Riverside, CA 92521, USA}

\begin{abstract}
We present Keck/MOSFIRE observations of the role of
environment in the formation of galaxies at $z\sim2$.  Using $K$-band
spectroscopy of H$\alpha$ and [\ion{N}{2}] emission lines, we have analyzed
the metallicities of galaxies within and around a $z=2.3$ protocluster
discovered in the HS1700+643 field.  Our main sample consists
of 23 protocluster and 20 field galaxies with estimates of
stellar masses and gas-phase metallicities based on the $N2$ strong-line metallicity indicator.
With these data we have examined the mass-metallicity relation (MZR) with
respect to environment at $z\sim2$.  We find that field galaxies
follow the well-established trend between stellar mass and metallicity,
such that more massive galaxies have larger metallicities.  The
protocluster galaxies, however, do not exhibit a dependence of metallicity
on mass, with the low-mass protocluster galaxies showing an enhancement in
metallicity compared to field galaxies spanning the same mass range.
 A comparison with galaxy formation models suggests that the mass-dependent
environmental trend we observed can be qualitatively explained in the context
of the recycling of ``momentum-driven" galaxy wind material.  Accordingly, winds
are recycled on a shorter timescale in denser environments, leading
to an enhancement in metallicity at fixed mass for all but the most
massive galaxies.  Future hydrodynamical simulations of $z\sim 2$
overdensities matching the one in the HS1700 field will
be crucial for understanding the origin of the observed environmental
trend in detail.         
\end{abstract}


\section{Introduction}

Galaxy clusters constitute the most massive gravitationally bound 
structures in the universe, making them excellent laboratories for 
studying galaxy formation in the most extreme environments.  Galaxy 
properties have been demonstrated to correlate with environment in the 
local universe \citep[e.g.,][]{hogg2004, 
kauffmann2004,clemens2006,vanderwel2010,wetzel2012}.  Particularly, galaxies found in 
the central, densest regions of clusters at $z<1$ are typically 
early-type galaxies with low star-formation rates (SFRs), while more 
active, late-type galaxies are preferentially located in the sparser 
edges of the cluster field.  These observations suggest that environment 
plays a key role in the evolution of galaxies, with large-scale 
overdense regions appearing to foster the earliest sites of galaxy 
formation.
 
Detailed cluster studies have been conducted out to $z\sim1$ 
\citep[e.g.,][]{holden2009,patel2009}, but there is limited information 
for clusters at $z>1$ \citep[but see e.g.,][]{papovich2012}.  Protoclusters at $z>2$ are 
still in the process of forming and have not yet virialized, making them 
perfect for studying the origin of the environmental trends observed 
locally.  Measuring the physical properties (e.g., stellar mass, 
metallicity, velocity dispersion) of protocluster galaxies will 
facilitate a better understanding of the processes that are relevant in 
dense environments and the formation times of cluster versus field 
galaxies, where field galaxies are defined as those not residing within an 
overdense region.

The metallicity of a galaxy reflects the past history of star formation, 
and is additionally modulated by gas inflows and outflows 
\citep{finlator2008,peeples2011,dave2012}.  A relationship between 
stellar mass and metallicity has been well established locally, where 
higher-mass galaxies have larger gas-phase metallicities 
\citep{tremonti2004}.  This trend has been shown to hold to at least 
$z\sim3$ \citep{erb2006a,maiolino2008}, though it appears to shift towards
lower metallicity at fixed mass as redshift increases.  Low-redshift studies of the 
mass-metallicity relation (MZR) with respect to environment have 
demonstrated that galaxies found in overdense regions have, on 
average, higher metallicities compared to field galaxies of the same 
mass \citep{cooper2008, ellison2009}. Similar measurements of metallicity
as a function of environment have yet to be performed at high redshift.

The newly commissioned Multi-Object Spectrometer For Infrared 
Exploration (MOSFIRE) instrument \citep{mclean2012} at Keck Observatory 
is ideal for studying high-redshift protoclusters due to its large field 
of view and multiplexing capabilities.  We have used MOSFIRE to collect 
a sizable sample of galaxies with $K$-band spectroscopy both within and 
outside of a redshift-space overdensity at $\langle z \rangle =2.3$ 
discovered in the HS1700+643 field \citep{steidel2005}.  Environmental
effects have already been detected in the HS1700 protocluster, such that
protocluster galaxies appear to have average ages and stellar masses
that are twice as large as those of galaxies in the surrounding, field
environment. Here we consider the connections between metallicity and
environment in the HS1700 field. For our sample of 
protocluster and field galaxies at $z\sim2$, we have measured the 
important rest-frame optical diagnostic emission lines H$\alpha$ and 
[\ion{N}{2}] to estimate the gas-phase oxygen abundance with the $N2$ strong-line metallicity 
indicator.    
This dataset will make possible for the first time the examination of the MZR for 
high-redshift protocluster galaxies compared to a field control sample, 
which will lead to a better understanding of how gas supplies are 
regulated during the formation of galaxies in extreme environments.

This paper is organized as follows.  In Section \ref{sec:samp} we 
discuss our sample selection.  The observations used for this study are 
presented in Section \ref{sec:obsdata}.  Section \ref{sec:phys} 
describes the physical properties of the galaxies measured for this 
analysis.  The MZR for protocluster galaxies is 
discussed in Section \ref{sec:mass_z}.  In Section \ref{sec:dis}, we examine various physical processes that can explain our observational results.
In Section \ref{sec:sum}, we summarize our findings and discuss their 
implications. A flat $\Lambda$CDM cosmology with $\Omega_{m}=0.3$, 
$\Omega_{\Lambda}=0.7$, and $H_{0}=70$ km s$^{-1}$ Mpc$^{-1}$ is assumed 
throughout.

\section{Sample Selection}
\label{sec:samp}

The HS1700+643 field ($\alpha=$ 17:01:01, $\delta=$ +64:11:58) was 
observed as part of the Keck Baryonic Structure Survey \citep[KBSS;][]{rudie2012}.  The KBSS was designed to explore the properties
of star-forming galaxies at $z\sim 2-3$ and the connection between
galaxies and the intergalactic medium (IGM) within the same cosmic volumes. 
As such, galaxies were targeted for spectroscopy in the fields of background, hyper-luminous
QSOs with high-resolution Ly$\alpha$ forest spectra.
The galaxy spectroscopic survey was conducted with the blue arm of the Low Resolution Imaging Spectrometer 
(LRIS-B) on the Keck I telescope \citep{oke1995,steidel2004}.  Galaxies 
were selected based on their rest-frame UV colors using the optical 
photometric $``$BX" ($z=2.20\pm0.32$) and $``$MD" ($z=2.73\pm0.27$) criteria of 
\citet{adelberger2004} and \citet{steidel2004,steidel2003}.
\citet{steidel2005} identified 
a highly significant ($\delta\sim7$) redshift-space overdensity of galaxies at $z=2.300$ in the $15${\farcm}$3\times15${\farcm}$3$ HS1700 
field in the course of the KBSS.  The majority of high-redshift protoclusters have
been identified around radio galaxies using narrowband 
imaging tuned to find Lyman Alpha Emitters (LAEs) 
\citep[e.g.,][]{venemans2007, kodama2007}. In contrast,
the HS1700 protocluster consists of a serendipitous discovery of a redshift spike
within a spectroscopic survey with a well-defined redshift selection function,
allowing for a robust estimate of galaxy overdensity.

We selected roughly equal numbers of UV spectroscopically-confirmed galaxies
within and outside the redshift-space overdensity 
in order to conduct a detailed analysis of how environment 
affects the physical properties of $z\sim2$ galaxies.  We define 
protocluster members as galaxies at $z=2.3\pm0.015$, while ``field" 
galaxies are defined as residing outside of the redshift-space overdensity, but between $2.0\lesssim z \lesssim$ 2.6.  

We used the MAGMA slitmask design tool\footnote{http://www2.keck.hawaii.edu/inst/mosfire/magma.html} to 
create three separate masks, which included 24 protocluster
galaxies and 21 field galaxies.  A sample of 46 photometric candidates 
without spectroscopic redshifts were included to fill any empty slits on 
our three MOSFIRE masks. The stellar mass (calculated from SED fitting,
see Section~\ref{sec:stelmass}) and $K_{s}$-band magnitude distributions
of our selected galaxies indicate that our MOSFIRE protocluster and field samples
are representative of their respective parent UV spectroscopic protocluster
and field samples.
For protocluster galaxies the 
mean and standard deviation for stellar mass and $K_{s}$ magnitude (AB) of the MOSFIRE and
parent samples are, respectively, $\langle \mathrm{log}(M_{\ast})_{\mathrm{MOS}} \rangle 
= 10.3 ~\mathrm{M}_{\odot} \pm 0.5$ and $\langle \mathrm{log}(M_{\ast})_{\mathrm{parent}} 
\rangle = 10.3~ \mathrm{M}_{\odot} \pm 0.5$, $\langle K_{s,\mathrm{MOS}} \rangle = 
22.6 \pm 0.5$ and $\langle K_{s,\mathrm{parent}} \rangle = 22.7 \pm 0.6$.  For field 
galaxies the mean stellar mass and $K_{s}$ magnitude (AB) of the MOSFIRE 
and parent samples are, respectively, $\langle \mathrm{log}(M_{\ast})_{\mathrm{MOS}} 
\rangle = 10.1 ~\mathrm{M}_{\odot} \pm 0.4$ and $\langle 
\mathrm{log}(M_{\ast})_{\mathrm{parent}}  \rangle = 10.2~\mathrm{M}_{\odot} \pm 0.5$, 
$\langle K_{s, \mathrm{MOS}} \rangle = 23.0 \pm 0.5$ and $\langle K_{s, \mathrm{parent}} \rangle = 22.9 
\pm 0.7$.

\section{Observations and Data Reduction}
\label{sec:obsdata}

Both photometric and spectroscopic data are required for the analysis of 
the MZR.  We utilized broadband ancillary photometry for our stellar 
mass estimates and MOSFIRE $\it{K}$-band spectroscopy for gas-phase 
metallicity.  In this section we describe both datasets and the 
reduction processes.

\subsection{Broadband Ancillary Photometry}
\label{sec:photo}

The HS1700 field has photometric data spanning from rest-frame 
ultraviolet to infrared wavelengths.  $U_{n}G\cal{R}$ imaging was 
performed on the William Herschel 4.2 m telescope (WHT) with the Prime 
Focus Imager.  Near-IR $JK_{s}$ imaging was obtained on the Palomar 5.1 
m Hale telescope with the Wide Field Infrared Camera (WIRC) and $H$-band 
imaging was completed with $HST$/WFC3-F160W.  Furthermore, HS1700 has 
photometric data from $\it{Spitzer}$/IRAC and MIPS.
A complete description of the reduction of these 
data is contained in \citet{reddy2012} and references therein.

\subsection{MOSFIRE Spectroscopy}

$\it K$-band near-IR spectroscopy of the HS1700 field was obtained using 
the MOSFIRE instrument on the Keck I telescope.  The 
MOSFIRE $\it K$-band filter is centered at 2.162 $\mu$m with a 
full-width at half-maximum (FWHM) of 0.483 $\mu$m.  The data were 
acquired over six nights during MOSFIRE commissioning runs on 2012 May 
6th, 7th, and 9th and June 2nd, 5th, and 6th.  As discussed in section 
\ref{sec:samp}, three separate MOSFIRE $\it{K}$-band masks were observed 
in the HS1700 field.  Each individual exposure time was 3 minutes, with a 
total integration time of 172 minutes, 180 minutes, and 45 minutes on, 
respectively, masks 1, 2 and 3.  All targets were observed with a slit 
width of 0${\farcs}$7 using an AB dither sequence with 3$^{\prime\prime}$ offsets.  The spectral resolution as determined 
from sky lines is $\sim$6 \AA, which corresponds to $\Delta\upsilon\sim$ 
80 km s$^{-1}$ ($R\simeq$3600).  The seeing FWHM over the six 
nights averaged $\sim$0${\farcs}$5.  We detected H$\alpha$ for 22/24 protocluster galaxies (92\%) and 15/17 field galaxies (88\%) for which H$\alpha$ fell within the MOSFIRE spectral format.\footnote{Of the original 21 targeted field galaxies, two galaxies were determined from H$\alpha$ to reside outside of the ``field" redshift space and two had redshifts such that H$\alpha$ fell off of the spectral format. Excluding these four objects results in a total of 17 field galaxies that were expected to have H$\alpha$ detections within the MOSFIRE $K$-band spectral format.}

There are 6 new protocluster galaxy detections, including BX759, which 
was previously identified as a field galaxy due to an incorrect LRIS redshift measurement and 5 galaxies without previously determined redshifts.  In addition there are 5 new field galaxy 
detections without previously determined redshifts.  We measured H$\alpha$ for a total of 28 protocluster 
galaxies and 20 field galaxies, of which 17 protocluster galaxies and 5 field galaxies also have [\ion{N}{2}] 
detections.  Protocluster member BNB19 was discovered using a 
narrowband filter tuned to detect Ly$\alpha$ emission.  
As all other objects in the sample were selected on the basis of their
rest-frame UV continuum colors, we  have excluded BNB19 from this analysis, resulting in
a final protocluster sample of 27 galaxies.

\begin{figure}[t]
\begin{center}
\centerline{
   \mbox{\includegraphics[scale =0.425]{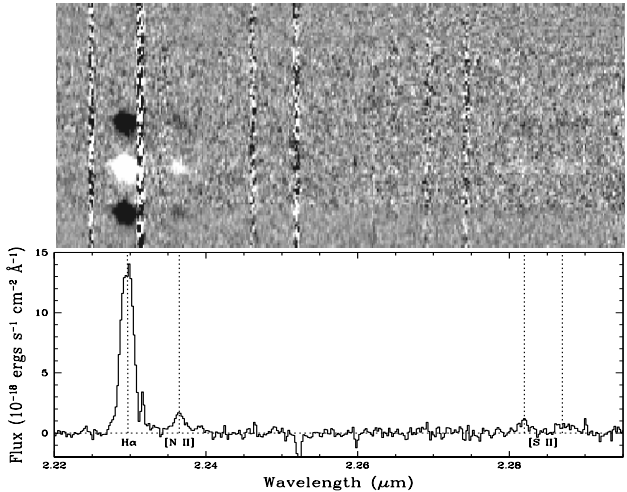}}
  }
\caption{(Top) Example MOSFIRE 2D spectrum for object BX490 produced 
from the MOSFIRE Data Reduction Pipeline.  (Bottom) Corresponding 1D 
spectrum after extraction and flux calibration.  In both spectra, 
H$\alpha$, [\ion{N}{2}]$\lambda$6584, and 
[\ion{S}{2}]$\lambda\lambda$6717,6732 are evident from left to right.
\label{fig:spectra}}
\end{center}
\end{figure}

We utilized the MOSFIRE data reduction pipeline (DRP)\footnote{http://code.google.com/p/mosfire/} to reduce all of the data.  
The DRP produces two-dimensional wavelength-calibrated, sky-subtracted, 
registered and combined spectra for each object, along with a 
corresponding inverse variance spectrum.  We then extracted one-dimensional spectra 
from the two-dimensional reduced science and variance images using the 
$\mathtt{IRAF}$ procedure {\tt{apall}}.  This routine summed the unweighted fluxes of pixels
inside of each extraction aperture.  The average aperture size along the slit was 1${\farcs}$4, with a range of $0{\farcs}9-2{\farcs}2$.  The resulting science and error spectra were placed on an absolute flux scale using observations of the A0 
star HD191225 obtained during the MOSFIRE runs.  Each flux-calibrated, 
one-dimensional spectrum was then placed in a vacuum, heliocentric frame 
(Figure \ref{fig:spectra}).

\section{Physical Properties}
\label{sec:phys}

\begin{deluxetable*}{lccccccr}
\tablewidth{0pt} \tabletypesize{\footnotesize}
\tablecaption{Physical Properties \label{tab:phys}}
\tablehead{
\colhead{Object} &
\colhead{$z_{{\rm H}\alpha}$} &
\colhead{~$\cal{R}$$^{\rm a}$~} &
\colhead{~$\it{K_{s}}$$^{\rm a,b}$~} &
\colhead{$M_{\ast}$$^{\rm c}$} &
\colhead{$F_{{\rm H}\alpha}$$^{\rm d}$ } &
\colhead{$F_{{\rm [NII]}}$$^{\rm d}$ } &
\colhead{12 + log(O/H)$^{\rm e}$}
}
\startdata
\\
\multicolumn{8}{c}{Protocluster Galaxies} \\
\\
BNB19$^{\rm f}$ & 2.2850 & 23.33 & -- & -- &  6.7$\pm$0.6 & 1.2$\pm$0.2 & 8.48$\pm$0.02\\
BM658 & 2.2939 & 25.29 &  -- & 0.3$\pm$0.1 & 1.0$\pm$0.1 & $<$0.1  & $<$8.37 \\
BX505$^{\rm g}$ & 2.3094 & 25.17 & 22.70 & 4.4$\pm$1.5 & 4.9$\pm$0.2 & 2.3$\pm$0.3 & 8.71$\pm$0.03 \\
BX563 & 2.2924 & 23.82 & 22.29 & 2.0$\pm$0.6 & 11.1$\pm$0.2 & 1.0$\pm$0.2 & 8.30$\pm$0.01\\
BX585 & 2.3085 & 24.71 & 23.35 & 0.5$\pm$0.2 & 3.9$\pm$0.2 & $<$0.2  & $<$8.18\\
BX649 & 2.2960 & 24.90 & 23.30 & 1.8$\pm$0.5 & 5.6$\pm$0.3 & 1.1$\pm$0.2 & 8.49$\pm$0.01\\
BX684$^{\rm f}$ & 2.2935 & 23.51  & -- & -- & 7.0$\pm$0.3  & 1.1$\pm$0.3 & 8.44$\pm$0.02\\ 
BX689$^{\rm f}$ & 2.2840 &  23.90  & --  & -- & 8.4$\pm$0.5 & 1.1$\pm$0.2 & 8.39$\pm$0.01\\ 
BX710 & 2.2960 & 23.96 & 22.03 & 3.6$\pm$0.7 & 15.8$\pm$0.2 & 1.7$\pm$0.2 & 8.35$\pm$0.01\\
BX711 & 2.2962 & 25.07 & 22.68 & 6.3$\pm$2.0 & 11.0$\pm$0.3 & 0.8$\pm$0.2 & 8.25$\pm$0.01\\
BX759 & 2.3096 & 24.43 & 23.08 & 1.3$\pm$0.6 & 2.2$\pm$0.3 & $<$0.3  & $<$8.43\\
BX763 & 2.2932 & 24.25 & 22.79 & 1.5$\pm$0.4 & 7.1$\pm$0.2 & 1.1$\pm$0.2 & 8.44$\pm$0.01\\
BX789$^{\rm f}$ & 2.3006 & 25.16 & -- & -- & 3.1$\pm$0.4 & $<$0.4 & $<$8.36\\
BX810 & 2.2937 & 24.68 & 22.61 & 2.3$\pm$0.8 & 9.9$\pm$0.4 & 1.5$\pm$0.3 & 8.43$\pm$0.01\\
BX879 & 2.3077 & 23.50 & 22.38 & 1.2$\pm$0.2 & 3.7$\pm$0.3 & 0.7$\pm$0.1 & 8.50$\pm$0.02\\
BX893 & 2.3092 & 24.38 & 23.20 & 2.7$\pm$1.3 & 1.7$\pm$0.3 & $<$0.3  & $<$8.46 \\
BX909 & 2.2949 & 23.73 & 22.27 & 3.7$\pm$0.9 & 14.8$\pm$0.4 & 1.6$\pm$0.3 & 8.35$\pm$0.01\\
BX913 & 2.2918 & 23.88 & 22.41 & 4.3$\pm$1.9 & 4.4$\pm$0.2 & 0.6$\pm$0.1 & 8.39$\pm$0.01\\
BX917 & 2.3081 & 24.43 & 21.88 & 7.6$\pm$1.4 & 14.0$\pm$0.5 & 2.1$\pm$0.3 & 8.43$\pm$0.01\\
BX918 & 2.3078 & 24.34 & 22.89 & 1.8$\pm$0.5 & 5.0$\pm$0.4 & $<$0.7  & $<$8.42\\
BX929 & 2.3081 & 23.83 & 22.88 & 1.9$\pm$0.4 & 5.0$\pm$0.5 & $<$0.3  & $<$8.18\\
BX935$^{\rm f}$ & 2.2860 & 25.18 & -- & -- & 4.2$\pm$0.4 & $<$0.3 & $<$8.27\\
BX939 & 2.2984 & 24.46 & 22.92 & 0.6$\pm$0.2 & 7.9$\pm$0.4 & $<$0.4  & $<$8.16\\
BX950 & 2.2968 & 24.51 & 22.51 & 1.5$\pm$0.6 & 3.3$\pm$0.3 & $<$0.5  & $<$8.43\\
BX951 & 2.3067 & 23.17 & 21.89 & 2.7$\pm0.7$ & 5.4$\pm$0.3 & 1.1$\pm$0.3 & 8.51$\pm$0.02\\
BX984 & 2.2976 & 23.51 & 21.97 & 2.3$\pm$0.6 & 10.3$\pm$0.4 & 1.1$\pm$0.3 & 8.34$\pm$0.01\\
MD109 & 2.2950 & 25.46 & 23.62 & 1.6$\pm$0.4 & 2.5$\pm$0.2 & $<$0.2  & $<$8.32\\
MD69 & 2.2899 & 24.85 & 21.90 & 20.1$\pm$1.4 & 9.4$\pm$0.4 & 2.8$\pm$0.2 & 8.60$\pm$0.01\\
\\
\hline
\\
\multicolumn{8}{c}{Field Galaxies} \\
\\
BM568 & 2.3901 & 24.72 & -- & 1.1$\pm$0.3 & 2.1$\pm$0.2 & $<$0.3  & $<$8.41\\
BM619 & 2.2668 & 24.04 & -- & 1.1$\pm$0.4 & 7.4$\pm$0.2 & $<$0.8  & $<$8.34\\ 
BX490 & 2.3973 & 22.88 & 21.84 & 1.2$\pm$0.1 & 30.2$\pm$0.4 & 3.4$\pm$0.2 & 8.36$\pm$0.01\\
BX535 & 2.6382 & 25.16 & 23.46 & 2.9$\pm$1.0 & 4.4$\pm$0.5 & $<$0.3  & $<$8.27\\
BX575 & 2.4348 & 23.82 & 22.74 & 0.6$\pm$0.2 & 8.8$\pm$1.1 & $<$0.6  & $<$8.22\\
BX592 & 2.4746 & 24.87 & -- & 0.8$\pm$0.3 & 2.2$\pm$0.2 & $<$0.2  & $<$8.33\\
BX604 & 2.2012 & 24.72 & 23.17 & 1.1$\pm$0.3 & 4.6$\pm$0.2 & $<$0.4  & $<$8.28\\
BX609 & 2.5714 & 24.12 & 23.20 & 0.8$\pm$0.1 & 6.2$\pm$0.5 & $<$0.4  & $<$8.23\\
BX625 & 2.0768 & 24.52 & 23.70 & 0.6$\pm$0.2 & 3.1$\pm$0.2 & 0.3$\pm$0.1 & 8.31$\pm$0.01\\
BX632 & 2.2353 & 25.12 & -- & 1.7$\pm$0.8 & 1.8$\pm$0.3 & $<$0.1  & $<$8.25\\
BX691 & 2.1912 & 25.33 & 22.53 & 9.9$\pm$1.6 & 9.6$\pm$0.3 & 1.9$\pm$0.3 & 8.50$\pm$0.01\\
BX708 & 2.4006 & 23.92 & 23.46 & 0.7$\pm$0.2 & 6.7$\pm$0.5 & $<$0.3  & $<$8.11\\
BX713 & 2.1394 & 24.48 & 23.03 & 4.7$\pm$1.8 & 4.9$\pm$0.2 & 0.4$\pm$0.2 & 8.27$\pm$0.02\\
BX717 & 2.4371 & 24.78 & 23.74 & 0.8$\pm$0.4 & 3.7$\pm$0.2 & $<$0.3  & $<$8.24\\
BX752 & 2.4016 & 24.86 & 22.69 & 4.4$\pm$1.0 & 8.7$\pm$0.7 & 2.4$\pm$0.3 & 8.58$\pm$0.02\\
BX772 & 2.3436 & 24.96 & 23.02 & 3.6$\pm$0.6 & 4.6$\pm$0.4 & $<$0.3  & $<$8.25\\
BX801 & 2.0393 & 24.11 & -- & 1.0$\pm$0.4 & 9.7$\pm$0.3 & $<$0.9  & $<$8.31\\
BX880 & 2.4381 & 24.98 & -- & 0.6$\pm$0.4 & 1.0$\pm$0.1 & $<$0.1  & $<$8.40\\
BX881 & 2.1822 & 24.99 & 23.95 & 0.4$\pm$0.2 & 1.3$\pm$0.2 & $<$0.7  & $<$8.76\\
MD77 & 2.5091 & 24.86 & 22.94 & 0.8$\pm$0.3 & 7.7$\pm$0.4 & $<$0.3  & $<$8.13\\
\enddata
\tablenotetext{a}{Magnitudes are on the AB system.}
\tablenotetext{b}{Objects with no reported $K_{s}$ magnitude are not detected down to a 3$\sigma$ limit of 24.05 AB.}
\tablenotetext{c}{Stellar mass in units of 10$^{10}$ M$_{\odot}$.}
\tablenotetext{d}{Emission-line flux and uncertainty in units of 10$^{-17}$ erg s$^{-1}$ cm$^{-2}$.}
\tablenotetext{e}{Errors listed are determined solely from propagation of the emission-line flux uncertainty measurements and do not include the systematic uncertainty in the $N2$ calibration.}
\tablenotetext{f}{Object not included in the MZR analysis.}
\tablenotetext{g}{Gas-phase oxygen abundance above solar.}
\end{deluxetable*}

\subsection{Stellar Masses}
\label{sec:stelmass}

We estimate the stellar masses for objects in our sample as in 
\citet{reddy2012}, using the full rest-frame UV through IR broadband spectral energy distribution (SED)
to find the best-fit stellar population model.  
A brief overview of the procedure is described here.

For stellar population modeling, we used the latest solar 
metallicity models of S. Charlot \& G. Bruzual (in preparation) which 
include the \citet{marigogirardi2007} prescription for the 
thermally-pulsating Asymptotic Giant Branch (TP-AGB) evolution of low- 
and intermediate-mass stars.  The broadband photometry was corrected in 
the optical bands for the effect of Ly$\alpha$ emission/absorption when 
the measurements were available.  The $K_{s}$ band was corrected for 
H$\alpha$ nebular emission as measured from our MOSFIRE spectra.  For 
each galaxy a constant, exponentially decreasing, and exponentially 
increasing star-formation history (SFH) was considered.  The best-fit 
model was determined by minimizing $\chi^{2}$ with respect to the 
observed photometry, yielding estimates of the SFR, age, $E(B-V)$, and 
stellar mass.  We assume a constant SFH with a minimum allowed age of 50 
Myr for all of our objects.  Using an exponentially decreasing or 
increasing SFH does not change our conclusions.

Stellar masses were estimated for 23 protocluster galaxies and 
all 20 field galaxies using the SED fitting procedure described above.  
Four protocluster galaxies (BX684, BX689, BX789, and BX935) do
not have sufficient multi-wavelength broadband photometric coverage 
for accurate stellar mass estimates, and are not 
included in the following mass-metallicity analysis.  
The average stellar mass estimated for 
protocluster members, $M_{\ast}= 3.2\times10^{10}~{\mathrm M}_{\odot}$, 
is approximately twice as large as that of the field galaxy sample, 
$M_{\ast}= 1.9\times10^{10}~{\mathrm M}_{\odot}$, which suggests that 
the protocluster galaxies have experienced a more advanced build-up in
stellar mass compared to field galaxies. This mass difference 
between protocluster and field galaxies was already noted by 
\citet{steidel2005}.

{The stellar mass error for each object was determined by a Monte Carlo
technique. Accordingly, each photometric data point was perturbed by a value
randomly drawn from a normal Gaussian distribution, the width of which was
set by the photometric error.  We performed 100 trials for each object.
In each trial, the best-fit SED was determined from the perturbed
photometry using the same method as for the actual photometry.
The stellar mass uncertainty was set equal to the standard
deviation of the distribution of stellar mass measurements
estimated from the artificial SEDs, utilizing a $3 \sigma$
clipping method to suppress the effect of outliers. Stellar masses and
corresponding errors are listed in Table \ref{tab:phys}.

\subsection{Metallicities}
\label{sec:metal}

Gas-phase metallicity reflects the entire past history of star formation, in 
addition to being affected by infall of metal-poor gas and outflow of 
metal-enriched gas \citep{finlator2008, dave2011, dave2012}.  There are 
several different methods typically used for evaluating the metallicity 
in a galaxy.   The most direct method employs the ratio between the 
[\ion{O}{3}]$\lambda$4363 auroral line and lower excitation lines such as [\ion{O}{3}]$\lambda\lambda$5007,4959, 
which allows a direct measurement of the oxygen abundance via the 
electron temperature ($T_{e}$) of the gas.  However, [\ion{O}{3}]$\lambda$4363 is typically too weak to detect in high-redshift galaxies.  Due to the difficulties in measuring the gas-phase 
metallicity with the direct method, empirical calibrations have been 
developed which fit a relationship between the direct $T_{e}$ 
method and strong-line ratios in H II regions.

To calculate the metallicity, we used the $N2$ indicator, $N2 \equiv$ 
log([\ion{N}{2}]/H$\alpha$), which traces
gas-phase oxygen abundance \citep{pettinipagel2004}. 
Since H$\alpha$ and [\ion{N}{2}] are close in wavelength, 
systematic uncertainties in $N2$ from dust extinction, flux 
calibration, and instrumental effects are insignificant.  However, one drawback of
$N2$ is the saturation of this ratio above solar metallicity, making it 
an unreliable tracer of metallicity in this regime.  To calculate $N2$, 
we measured the flux for H$\alpha$ and [\ion{N}{2}] simultaneously using 
a fixed central wavelength and FWHM based on the best-fit parameters for H$\alpha$, which 
is the brighter line.  A combined fit was obtained for both emission 
lines by fitting Gaussian profiles using the $\tt{IRAF}$ task, 
$\tt{splot}$.

We employed a Monte Carlo technique to measure the uncertainties in the emission line 
centroid, flux, and FWHM.  For each object, 500 fake spectra were 
created by perturbing the flux at each wavelength of the true spectrum 
by a Gaussian random number with the standard deviation set by the level 
of the $1 \sigma$ error spectrum.  Line measurements were obtained from 
the fake spectra in the same manner as the actual data. The standard 
deviation of the distribution of measurements from the artificial 
spectra was adopted as the error on the centroid, flux, and FWHM.  For 
objects with no detected [\ion{N}{2}] emission, 500 simulated spectra 
were created in the same manner as described above to measure upper 
limits.  The flux was summed at the nominal position of [\ion{N}{2}] 
over an interval defined by the extent of the measured 
H$\alpha$ emission line above the continuum in each simulated spectra.  We defined the $1 \sigma$ upper 
limit as the standard deviation of the distribution of the calculated 
flux from the artificial spectra.

To determine the gas-phase oxygen abundance from $N2$, we used the 
calibration from \citet{pettinipagel2004}, which was established from a 
linear fit to a large sample of local extragalactic H II regions with 
both $N2$ measurements and direct $T_{e}$-based oxygen abundance 
estimates. According to this calibration,

\begin{equation}
12 + {\rm log(O/H)} = 8.90 + 0.57 \times N 2 
\end{equation}

\noindent This relation has an inherent $1 \sigma$ dispersion of 
$\pm0.18$ dex, which is the dominant source of error in our metallicity 
measurements for individual galaxies.  It should be noted that since the conversion between $N2$ and gas-phase oxygen abundance was calibrated from a sample of low-redshift galaxies, this conversion may not be valid for high-redshift galaxies \citep{shapley2005,liu2008}. 
For example, unknown differences in HII-region physical conditions may pertain at high redshift, extending to the electron density, ionization parameter and nitrogen-to-oxygen abundance. These evolutionary differences may lead to both a possible systematic offset and increased scatter in the N2 indicator.
 However, the differential nature of this study between protocluster and field galaxies at $z\sim2$ mitigates the importance of a possible divergence between low- and high-redshift objects.  The uncertainty in the calibration, though, may result in a $1 \sigma$ dispersion larger than 0.18 dex at higher redshift. 
Metallicities and uncertainties are listed for each object in Table \ref{tab:phys}.  The metallicity uncertainties stated in Table \ref{tab:phys} were estimated solely from propagation of the emission-line flux errors as calculated from the Monte Carlo method described above.  These errors do not include the systematic uncertainty in the $N2$ calibration. 

\section{Mass-Metallicity Relation}
\label{sec:mass_z}

The existence of a correlation between the mass of a galaxy and 
the gas-phase metallicity was first noted in the late 1970s by 
\citet{lequeux1979}.  Since then the correlation between metallicity and 
stellar mass has been firmly established for nearby galaxies and traced out to $z\sim3$ 
for star-forming galaxies \citep{tremonti2004,erb2006a,kewley2008, 
mannucci2009}.  The source of this relation, however, remains uncertain.  
Numerous models have been created to explain the origin and observed 
evolution of the MZR \citep[e.g.,][]{finlator2008, oppenheimer2008, 
dave2012}.  Observations of the MZR provide the much needed constraints 
for these galaxy formation models in order to understand how the gas 
supply is regulated in galaxies.

At $z\sim0$ it has been demonstrated that galaxies found in overdense 
regions have enhanced gas-phase metallicities compared to field galaxies 
at the same stellar mass \citep{cooper2008, ellison2009}.  At higher 
redshift, the relationship among mass, metallicity, and environment has not been well 
studied due to small sample sizes and a lack of robust environmental measurements.  Examining the MZR with respect to 
environment provides a unique probe into how baryons cycle in and out of 
galaxies, and offers additional constraints on galaxy formation models.  
Our measurements of gas-phase metallicities of galaxies within and surrounding
the HS1700 protocluster present a special opportunity for connecting metallicity
and environment at early times.

Figure \ref{fig:mz} shows the gas-phase metallicity versus stellar mass 
for protocluster (in red) and field galaxies (in black) 
in our sample.  Downward arrows indicate $1 \sigma$ upper limits.  The data 
from \citet{erb2006a} are also shown (in blue) in Figure \ref{fig:mz}.  
These authors constructed the MZR based on the $N2$ indicator for a sample of 87 
star-forming galaxies at $\langle z \rangle=2.3$ binned into composite spectra by stellar mass.
In contrast, the MOSFIRE data points are from 
$\emph{individual}$ objects.  Measurements of individual galaxies are important for understanding the inherent scatter of the MZR, in addition to enabling a more direct, object-by-object, comparison to other physical properties such as SFRs.  The dotted curve is the MZR derived by \citet{erb2006a} using  
the $N2$ indicator for $\sim$53,000 SDSS galaxies \citep{tremonti2004} 
after the application of a downward shift of 0.56 dex as implied by the \citealt{erb2006a} data. 
The protocluster and field galaxies are distributed around the fit, with 
the majority appearing to fall below the \citet{erb2006a} trend, which 
corresponds to a lower metallicity for a given stellar mass.  The dashed 
horizontal line indicates solar metallicity, where the $N2$ indicator 
saturates.  One of our objects, BX505, has a measured gas-phase 
metallicity above solar. 

\begin{figure}[h!]
\begin{center}
\centerline{
   \mbox{\includegraphics[scale =0.48]{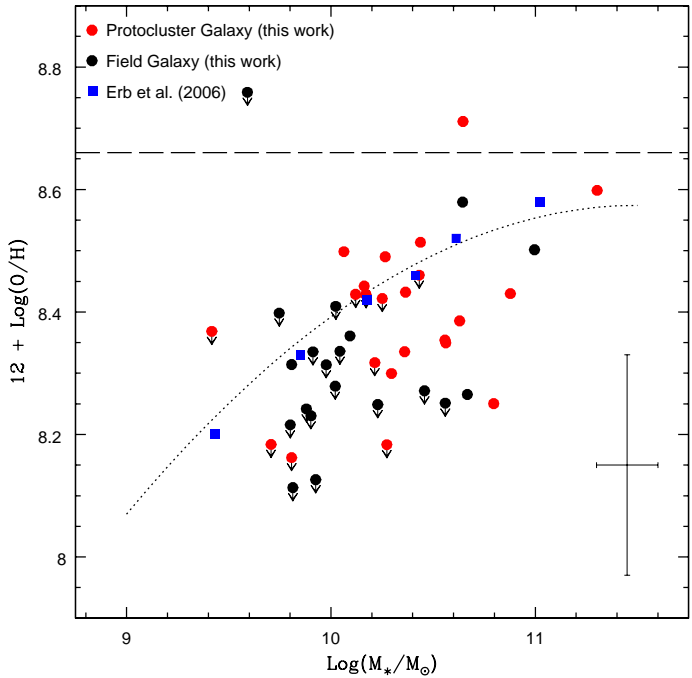}}
  }
\caption{MZR for the protocluster (red 
circles) and field (black circles) galaxy samples at $z\sim2$ using the 
$N2$ indicator.  Downward arrows represent $1\sigma$ upper limits in [\ion{N}{2}] to H$\alpha$.  The dashed horizontal line represents solar abundance, 
where $N2$ saturates.  The blue squares are from \citet{erb2006a}, indicating
star-forming galaxies at $\langle z \rangle=2.3$, based on composite 
spectra binned by stellar mass.  The dotted curve is the 
MZR based on $\sim$53,000 SDSS galaxies after a 
downward shift of 0.56 dex, as in \citet{erb2006a}.  Error bars are shown in the lower-right corner, indicating the average errors
in stellar mass and gas-phase metallicity. The metallicity error is
dominated by the calibration uncertainty from the $N2$ indicator.
\label{fig:mz}}
\end{center}
\end{figure}

To detect differences between the protocluster and field 
galaxy samples, we binned the galaxies by stellar mass and constructed
composite MOSFIRE spectra for each bin.   A caveat related to using composite spectra is that they do not reflect the scatter among the individual points.  However,  given that more than half of the objects in the sample have measured upper limits in [\ion{N}{2}]/H$\alpha$, composite spectra offer the most robust method to examine the average   
 [\ion{N}{2}]/H$\alpha$ ratio and corresponding metallicity in each bin.  We created
low- and high-mass bins for both protocluster and field samples.  These bins were designed
so that the average mass was roughly equivalent for the corresponding
protocluster and field subsamples at both low and high mass.  
For the protocluster sample, there are 6 galaxies in the lower mass bin
and 17 in the higher mass bin. For the field sample, there are 14 galaxies in the lower mass bin
and 6 in the higher mass bin.  To create the composite 
spectrum for each bin, we first shifted each individual spectrum into the rest frame using 
the redshift calculated from the observed H$\alpha$ wavelength.  Next we 
converted each spectrum from $F_{\lambda}$ to $L_{\lambda}$ using the
monochromatic luminosity distance to remove
any redshift dependence on the observed flux.  
Finally, the $\texttt{IRAF}$ task $\tt{scombine}$ was 
used to compute the average at each wavelength of the rest-frame spectra 
with a minimum/maximum pixel rejection 
(Figure \ref{fig:comp_spec}).  H$\alpha$ and  [\ion{N}{2}] were
measured and the corresponding $N2$-based metallicities estimated for each composite spectrum in 
the same manner as described for the individual galaxies.

\begin{figure}[t]
\begin{center}
\centerline{
   \mbox{\includegraphics[scale =0.45]{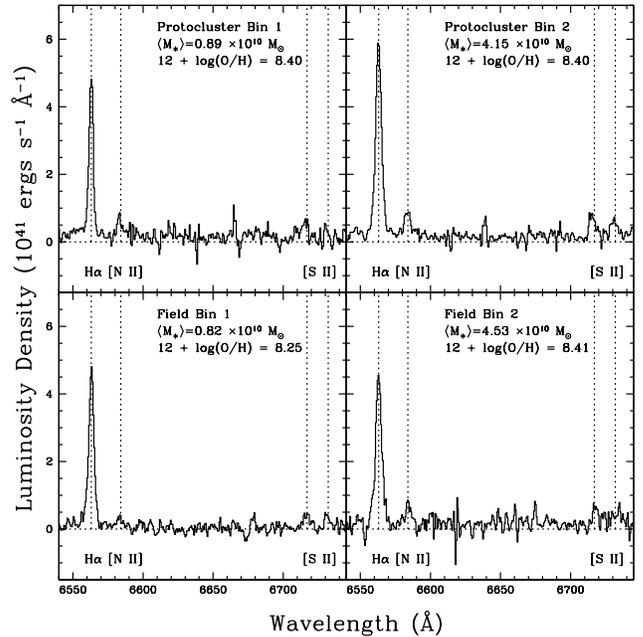}}
  }
\caption{Rest-frame composite spectra for each protocluster and field 
galaxy bin.  The first and second columns are, respectively, the low and 
high mass bins.  The top row features protocluster galaxy composites and the 
bottom row shows the field galaxy composites. From left to right, 
H$\alpha$, [\ion{N}{2}]$\lambda$6584, and 
[\ion{S}{2}]$\lambda\lambda$6717,6732 are marked on each spectrum.  In 
addition, the mean stellar mass and metallicity are 
labeled for each bin.
\label{fig:comp_spec}}
\end{center}
\end{figure}

Figure \ref{fig:mz_comp} shows the gas-phase oxygen abundance calculated 
with the $N2$ indicator from each mass bin plotted versus the corresponding 
mean stellar mass.  The horizontal bars show the range of stellar masses 
in each bin.  The metallicity measurement errors were calculated from
a Monte Carlo approach taking into account both
sample variance and stellar mass uncertainties.
We generated 500 realizations of the low- and high-mass
composite spectra for both the protocluster and 
field samples. In each realization, in order to account
for the error in stellar mass, the stellar mass of each galaxy was randomly perturbed according
to a Gaussian distribution with the width set by the 1$\sigma$
stellar mass error. Both samples of protocluster and field galaxies
were then sorted by the perturbed stellar mass, and divided into low-
and high-mass bins as defined by the mass ranges of the original
composite bins. Despite using perturbed masses, we found that the numbers
of objects in each mass bin were typically consistent with the numbers in the original,
unperturbed bins.  Finally, to account for sample variance,
the protocluster and field galaxies
in each mass bin were bootstrap resampled.  The vertical metallicity 
error bars include the error from the $N2$ calibration of 12 +  log(O/H), reduced by 
$N^{1/2}$ where $N$ is the number of objects in the original composite spectrum, 
as well as the uncertainties in the measurements of the H$\alpha$ and 
[\ion{N}{2}] fluxes from the bootstrap Monte Carlo method.  The field 
galaxies in Figure \ref{fig:mz_comp} show a clear trend similar to the one in
\citet{erb2006a}, but shifted down by $\sim 0.1$ dex in metallicity at fixed stellar mass.  
To quantify the significance of this difference, we applied
best-fit linear regressions to both the field galaxies and the 
\citet{erb2006a} data. The parameters of the linear regressions
to the \citet{erb2006a} and field galaxy datasets are consistent
within the $1 \sigma$ errors, demonstrating that the two datasets
are not significantly different.

\begin{figure}[t]
\begin{center}
\centerline{
   \mbox{\includegraphics[scale =0.45]{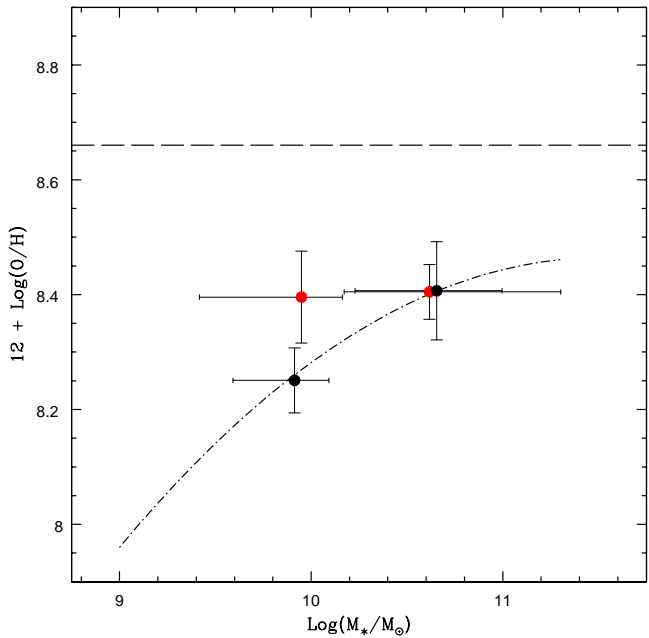}}
  }
\caption{MZR for protocluster and field galaxies 
binned by stellar mass.  The horizontal bars represent the mass range in 
each bin.  The vertical error bars represent the $1 \sigma$ error from the 
emission line measurements and the $N2$ calibration.  The dotted line is 
the same as from \citet{erb2006a} except that it has been shifted 
downward by 0.67 dex from the \citet{tremonti2004} 
fit instead of 0.56 dex.  The field galaxies appear to follow the trend presented in 
\citet{erb2006a}, while the protocluster galaxies show no strong 
correlation with mass and metallicity, with both protocluster galaxy 
mass bins having approximately the same measured gas-phase metallicity.
\label{fig:mz_comp}}
\end{center}
\end{figure}

Figure \ref{fig:mz_comp} also indicates that there is no 
strong correlation between mass and metallicity for protocluster 
galaxies, with both mass bins having approximately the same measured 
gas-phase metallicity.  
A comparison between protocluster and field trends reveals that
the high-mass protocluster and field
bins are consistent with each other in metallicity. It is at lower masses
that the two samples diverge, with the protocluster metallicity
notably higher than that of the field sample.

\section{Discussion}
\label{sec:dis}

In this section, we consider physical mechanisms that may give rise to the observed $z\sim2$ trends
between environment and metallicity in the HS1700 field.  \citet{dave2011} utilize
cosmological hydrodynamical simulations of galaxy formation to investigate how inflows, outflows,
and star formation affect the metal content and gas fraction in a galaxy over cosmic time.
Specifically, these authors examine the effect of local environment on the metallicities of
simulated galaxies. In \citet{dave2011} environment is defined by measuring the local galaxy
density in a 1$h^{-1}$ Mpc top-hat sphere.  Galaxies at densities greater than $0.5 \sigma$ above
the mean are considered to reside in high-density regions.  The \citet{dave2011} simulations
reveal an enhancement of the mean metallicity for objects found in overdense regions at $z=0$
using either ``momentum-driven winds" or ``no-winds" models. \citet{finlator2008} also
advance the model of momentum-driven winds to reproduce the MZR from
\citet{tremonti2004} at $z\sim0$ and \citet{erb2006a} at $z\sim2$.
Outflows have been shown to be ubiquitous for $z\sim 2-3$ UV-selected star-forming
galaxies \citep{pettini2001, shapley2003,steidel2010}.  Therefore, the no-winds model clearly does not
provide an accurate description of these
systems, leaving the momentum-driven scenario. With this prescription for winds,
the enhancement in metallicity disappears in galaxies at
masses greater than $10^{11}$ M$_{\odot}$ (see Figure 14 in \citealt{dave2011}).

\citet{steidel2005} established that the HS1700 protocluster corresponds to a physical overdensity
of ${\Delta \rho}/{\langle \rho \rangle} \sim2$. Qualitatively, the momentum-driven winds model can replicate the MZR of the
protocluster members if we assume that the \citet{dave2011} results for environmental dependence
exist at $z\sim2$ for an overdensity similar to that found in HS1700.  In detail,
\citeauthor{dave2011} use a different density estimator, which prevent us from making a quantitative comparison.  In addition, these authors indicate that the environmental metallicity
enhancement disappears at a stellar mass that is a factor of two more massive than suggested in
our results.  This discrepancy may be attributed to the difference in redshift between the
\citeauthor{dave2011} simulated objects at $z=0$ and our galaxies at $z\sim2$ if the transition
for the disappearance of the metallicity enhancement occurs at lower masses during earlier epochs.

\citet{dave2011} suggest that the metallicity enhancement in overdense environments at $z=0$ may
be due to the differentially higher enrichment of intergalactic gas in such regions,
resulting in cluster galaxies accreting additional metals compared with field galaxies.
At $z\sim2$, however, when the IGM has had less time for differential enrichment, the significance of
this effect is unclear. ÊFurthermore, the observed metallicity enhancement for HS1700 protocluster
galaxies exhibits a mass dependence, with a larger metallicity offset at lower stellar
masses.  This mass dependence indicates that a general enhancement of the metallicity of infalling gas in an overdensity
cannot alone explain the environmental MZR trend we observe.
\citet{dave2011} turn to the results of \citet{oppenheimer2008}
to explain the origin of a mass-dependent environmental trend in metallicity.

As reported by \citet{oppenheimer2008}, environment is a primary factor determining
how far winds can carry metals, and how long it will take for the metals to fall back into the galaxy
and be $``$recycled."  Specifically, these authors demonstrate in their simulation that the timescale for metals
to be recycled is inversely correlated with environmental density, such that metals are
recycled more rapidly in denser environments (see panel (b) of Figure 12
and Figure 14 in \citealt{oppenheimer2008}).
Accordingly, low- to intermediate-mass protocluster galaxies will recycle their metals
more efficiently due to residing in an overdense environment, compared with
field galaxies of the same mass.  However, at the highest masses, wind recycling is so
effective that all galaxies re-accrete their ejected metals
quickly regardless of environment.  These mass-dependent results of wind recycling are reflected
qualitatively in our Figure \ref{fig:mz_comp}, with an offset between the protocluster and field
MZRs at lower stellar mass, but not at higher stellar mass. In detail, however,
the actual high mass threshold at which the environmental
effect disappears must also be matched.  Models of galactic outflows at $z\sim 2$ including environments
similar to the HS1700 overdensity are needed to investigate if the environmental dependence
of wind recycling is a viable explanation for the MZR trends we observe.

\section{Summary}
\label{sec:sum}

In this paper we utilize Keck/MOSFIRE data to examine the gas-phase 
oxygen abundance of 23 protocluster and 20 field galaxies with respect to stellar mass in the HS1700 field.  Metallicities were measured from the [\ion{N}{2}]/H$\alpha$ ratio, while stellar masses were estimated from stellar population modeling of rest-frame UV through IR SEDs.   We first examined the 
correlation between stellar mass and metallicity for each $individual$ galaxy and 
conclude that the sample is distributed around the \citet{erb2006a} 
result, with the majority of galaxies found below this trend.  We then 
created composite spectra to study the MZR for protocluster and field galaxies separately.
The field composite datapoints are slightly lower in metallicity but consistent with previous studies of $z\sim2$ star-forming galaxies 
within the $1 \sigma$ errors.  The protocluster galaxies, however, demonstrate no 
correlation between metallicity and stellar mass.  In detail, the protocluster and field high-mass bins for both samples are consistent, while the low-mass protocluster bin is measured to have a higher metallicity than that of the corresponding low-mass field bin.

Recent cosmological hydrodynamical simulations of galaxy formation 
suggest that winds carrying metals will recycle more efficiently back to the host 
galaxy in an overdense environment, which may lead to cluster galaxies having enhanced gas-phase metallicity at moderate stellar mass.  At high 
stellar masses, though, wind recycling is so effective that all galaxies 
re-accrete their ejected metals quickly.  Galaxy formation models featuring an environment similar to that in HS1700 at $z\sim2$ need to be investigated to understand if environmental dependence of enriched infall caused by wind recycling provides a viable explanation for our 
observations.  This 
work represents the first analysis of an environmental dependence of the MZR at $z\sim2$, a crucial step towards 
understanding how gas reservoirs are regulated in extreme environments 
during galaxy formation.

\acknowledgments 
MOSFIRE was developed by the consortium of the University of California, Los Angeles, the California Institute of Technology, the University of California, Santa Cruz and the W. M. Keck Observatory.  Funding was provided by a grant from the National Science Foundation's Telescope System Instrumentation Program and by a generous donation from Gordon and Betty Moore to the Keck Observatory.  We would like to thank the entire MOSFIRE team, whose dedication and hard work made this project possible.  We also gratefully acknowledge the outstanding support of the entire Keck Observatory staff during commissioning, especially Marc Kassis, Greg Wirth and Al Honey.  A.E.S. acknowledges support from the David and Lucile Packard Foundation.  N.P.K. acknowledges support from the NSF grant 1106171.  Additionally, we would like to acknowledge Romeel Dav\'e for insightful conversations.  Finally, we wish to extend special thanks to those of Hawaiian ancestry on whose 
sacred mountain we are privileged to be guests.  Without their generous hospitality, most of the observations presented 
herein would not have been possible.

\bibliographystyle{apj}

\end{document}